# Whole counting vs. whole-normalized counting: A country level comparative study of internationally collaborated papers on Tribology

B. Elango[1*] and P. Rajendran[2]

[1]Library,
IFET College of Engineering,
Villupuram,
Tamilnadu,India.
Email: elangokb@yahoo.com

[2]University Library,
SRM University,
Kattangulathur,
Tamilnadu,India.

*Corresponding author

# Whole counting vs. whole-normalized counting: A country level comparative study of internationally collaborated papers on Tribology

**ABSTRACT**

The purpose of this study is to compare the changing behavior of two counting methods (whole counting and whole-normalized counting) and inflation rate at country level research productivity and impact. For this, publication data on tribology research published between 1998 and 2012 from SCOPUS has been used. Only internationally collaborated papers are considered for comparison between two counting methods. The result of correlation tests shows that there is highly correlation in all the four indicators between the two counting methods. However, the result of t-test shows that there is significant difference in the three indicators (paper count, citation count and h-index) between the two counting methods. This study concludes that whole-normalized counting (fractional) is the better choice for publication and citations counting at the country level assessment.

**INTRODUCTION**

The counting of co-authored publications constituted methodological problem in scientometrics based research evaluation (Huang and Lin 2011)[1]. The counting of papers and citations is fundamental, to the assessment of research productivity and impact. In the increasing trend of scientific collaboration across international borders, it becomes conceptually and methodologically challenging to conduct counting for an internationally collaborated paper (Huang, Lin and Chen 2011[2]; Lin, Huang and Chen 2013[3]). Egghe, Rousseau and van Hooydonk (2000)[4] concluded that counting methods may have important practical consequences for career opportunities. According to National Science Foundation

(www.nsf.gov/statistics/srs11201/content.cfm), there are two well-known counting methods in the evaluation studies such as whole counting and fractional counting. Leiden Ranking supports this two counting methods - full and fractional (Waltman, et al 2012)[5]. Chudlarský, Dvořák and Souček (2014)[6] considered the two counting methods namely, fractional and whole counting for the comparison of research output counting methods at the institutional level. A recent study by Aksnes, Schneider and Gunnarsson (2012)[7] has found that the difference between whole and fractionalized counts is generally greatest for the countries with the highest proportion of internationally co-authored articles. Whole counting and fractional counting are rank independent while straight counting is rank dependent (Zheng, et al 2014)[8]. Further they found that counting methods have minor effects on country rankings in patent count, citation count and CPP ratio. Analyses by Huang et al (2011)[2] and Lin et al (2013)[3] supported the use of straight counting and fractional counting methods at the country level assessment. Huang and Lin (2010)[9] found that all the five accounting procedures are capable of yielding reliable comparisons of national research productivity. Further they found that country ranks were not affected greatly by accounting procedures. Huang, Lin and Chen (2011)[2] found that the whole counting method gives each collaborating country one full credit may not be the best counting method. According to Gauffriau et al (2008)[10], fractional count is more logical than whole count. Sung et al (2014)[11] observed that whole count can be seen as an indicator for directly measuring a country's R&D output. The aim of this study is to compare the two counting methods (whole counting and whole-normalized counting) on country rankings based on the research productivity (paper counts) and research impact (citation counts, citations per paper and h-index) using the papers published on tribology between 1998 and 2012. Unlike the previous studies (e.g. Huang, Lin and Chen 2011[2]; Huang and Lin 2011[1]; Chudlarský, Dvořák and Souček 2014[6]), only internationally collaborated papers are considered for the comparison of ranking and inflation rate in this study. Further, among the various available h-index models, we use the Glanzel-Schubert model for comparison.

## METHODOLOGY

### Data Set

For this study, we have used the bibliographic data on tribology research indexed in SCOPUS. The following search query was used (Elango, Rajendran and Bornmann, submitted)[12]: *tribolog* OR "tribosyst*" OR "tribo-syst*" OR "tribo-chem*" OR "tribochem*" OR "tribotechn*" OR "tribo-physi*" OR "tribophysi*"*. Search was limited to Articles, Conference Papers and Reviews published in Journals indexed under Physical Sciences. The search yielded 27952 records for the period 1998-2012. The retrieved data has been extracted as CSV file. Of this, 3789 internationally collaborated papers have been considered for analysis. If a paper contains more than one country name in the affiliations field, then it is an internationally collaborated paper. Countries having at least 100 papers in the data setare considered for the comparisonof two counting methods (figure 1).

### Tools used

#### CPP

Citations per Paper (CPP) can be used to assess the impact of publications for publication years, countries, institutes and authors (Elango, Rajendran and Bornmann 2013)[13]. It is obtained by dividing the total number of citations by total number of papers.

#### h-index

h-index is an indicator that measures both the research productivity and impact of a researcher (Huang and Lin 2011)[1]. Since its introduction in 2005 by Hirsch[14], it has been applied not only to individual researchers, but also to countries (Schubert 2007)[15]. For example, Elango, Rajendran and Bornmann (2013)[13] applied the h-index to countries in the field of nanotribology. Among the various h-index models, the following model suggested by Glänzel and Schubert (2007)[16] has been used to assess the impact of countries.

$$h = cP^{1/3}(CPP)^{2/3}$$

Where, c is a constant (0.9 for journals and 1 for others), P is a number of papers and CPP is a Citations per Paper.

Counting methods used

Among the various available counting methods, the following two counting methods (OECD, 2009)[17] have been used in this study.

**Whole Counting** (WC) : Each unique collaborating country receives one full credit. It is also called as normal or standard counting. For example, an article with three addresses of which two are from France and one from Germany, both the countries receive one full credit.

**Whole-Normalized Counting** (WNC) : Each unique collaborating country receives equal share of the credit. It is also called as fractional counting. For example, an article with three addresses of which two are from France and one from Germany, attracts 1/2 credit for both the countries.

Inflation Rate

Inflation rate is obtained by dividing the paper count, citation count, CPP and h-index from whole counting by those from whole-normalized counting.

**Statistical procedures**

Pearson and Spearman's tests have been performed to determine the relationship that exists between the two counting methods. These tests are done in SPSS and MS-Excel. Further t-test ([www.socscistatistics.com](www.socscistatistics.com)) was also performed.

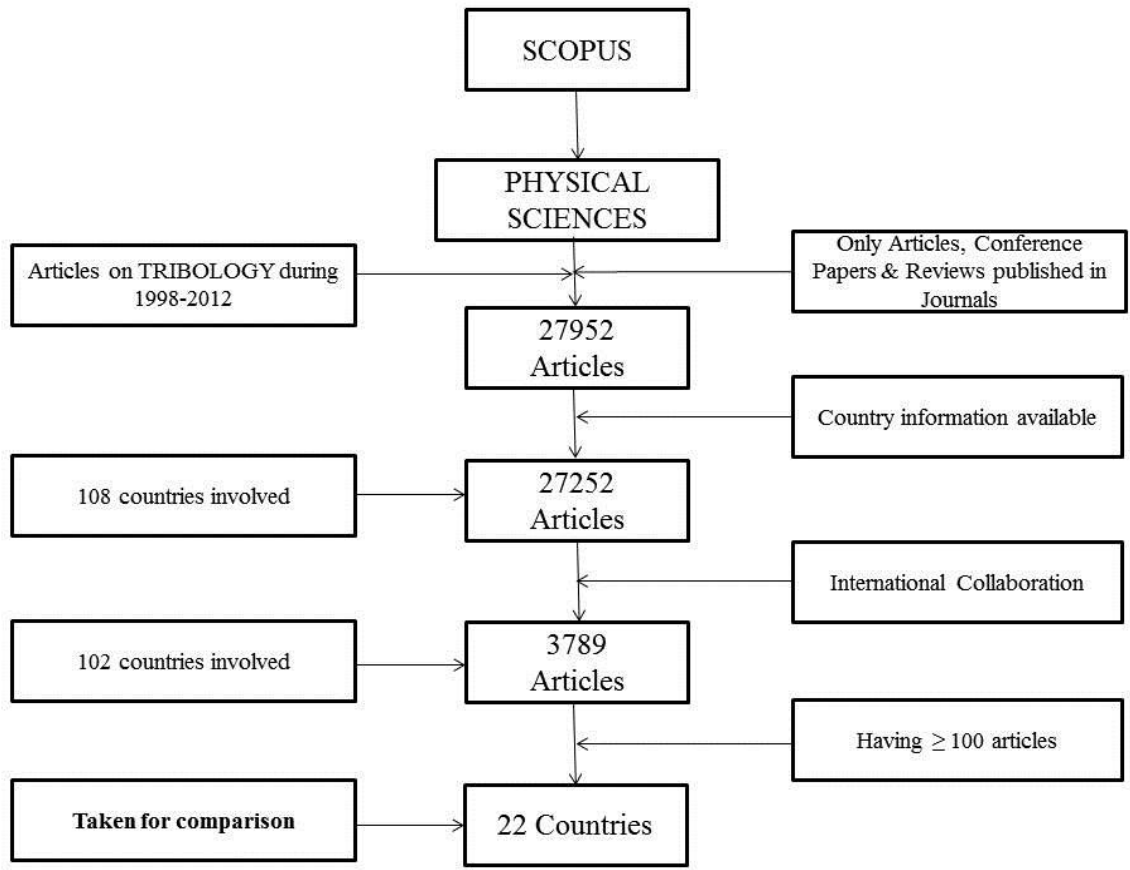

Figure 1 : Selection of dataset

**RESULTS**

A total of 27952 papers have been published on tribology during a span of 15 years from 1998 to 2012. Of 27952 papers, 27252 papers are having country information of the contributing authors. Out of 27252 papers, 3789 papers were contributed by authors from more than one country. These international collaborative papers accounted for almost 14%. It is very low compared to physics literature where it was 23% (Huang and Lin 2010)[9] during the period 1989-2008. These international collaborative papers received 44340 citations with an average of 11.70 citations per paper from its time of publication through the date of database access on 19.12.2013.

**Paper count and ranking of top countries**

Table 1 shows the paper counts and relative rank of top countries by the two counting methods. The ranks of top countries are similar by two counting methods except Hong Kong,

Brazil, Belgium and Sweden. A mutual exchange is observed between the pairs Hong Kong – Belgium and Sweden – Brazil. Inflation rate is calculated by dividing the paper count by whole counting with paper count by whole-normalized counting. It is observed that inflation rate of paper count from whole-normalized counting to whole counting for top countries ranges between 1.97 (Brazil) and 2.25 (Sweden). The range of inflation rate is contrast to previous studies in paper count where it was between 1.13 and 1.73 (Huang, Lin and Chen 2011)[2] and in patent count where it was between 1.01 and 1.38 (Zheng, et al 2014)[8].

| Table 1 – Paper counts and rankings | | | | | |
|---|---|---|---|---|---|
| Country | Paper Count | | Rank | | Inflation Rate (WC / WNC) |
| | WC | WNC | WC | WNC | |
| United States | 1098 | 519.58 | 1 | 1 | 2.11 |
| China | 719 | 348.27 | 2 | 2 | 2.06 |
| Germany | 594 | 277.28 | 3 | 3 | 2.14 |
| United Kingdom | 578 | 269.58 | 4 | 4 | 2.14 |
| France | 522 | 245 | 5 | 5 | 2.13 |
| Japan | 431 | 199.33 | 6 | 6 | 2.16 |
| Canada | 248 | 114.38 | 7 | 7 | 2.17 |
| Spain | 229 | 104.92 | 8 | 8 | 2.18 |
| South Korea | 204 | 98.08 | 9 | 9 | 2.08 |
| India | 201 | 95.88 | 10 | 10 | 2.10 |
| Poland | 188 | 87.25 | 11 | 11 | 2.15 |
| Italy | 185 | 84.7 | 12 | 12 | 2.18 |
| Switzerland | 182 | 82.4 | 13 | 13 | 2.21 |
| Russian Federation | 181 | 82.03 | 14 | 14 | 2.21 |
| Austria | 136 | 62.37 | 15 | 15 | 2.18 |
| Australia | 134 | 62.37 | 16 | 16 | 2.15 |
| **Hong Kong** | 121 | 58.5 | **19** | **17** | 2.07 |
| **Brazil** | 115 | 58.5 | **20** | **18** | 1.97 |
| **Belgium** | 126 | 57.17 | **17** | **19** | 2.20 |
| **Sweden** | 125 | 55.67 | **18** | **20** | 2.25 |
| Portugal | 112 | 50.75 | 21 | 21 | 2.21 |
| Netherlands | 107 | 49.28 | 22 | 22 | 2.17 |

**Citation count and ranking of top countries**

Table 2 shows the citation counts and relative rank of top countries by the two counting methods. The ranks of top countries are similar by both the methods except

Sweden, India, Poland, South Korea, Netherlands and Australia where there is a mutual exchange has been observed between two countries. Inflation rate is calculated by dividing the citation count by whole counting with citation count by whole-normalized counting. It is observed that inflation rate of citation count from whole-normalized counting to whole counting of top countries ranges between 2.04 (Hong Kong and Brazil) and 2.37 (Netherlands). The range of inflation rate is contrast to previous studies in paper count where it was between 1.24 and 2.16 (Huang, Lin and Chen 2011)[2] and in patent count it was between 1.02 and 1.44 (Zheng, et al 2014)[8].

| Table 2 – Citation counts and rankings ||||||
| Country | Citation count || Rank || Inflation Rate (WC / WNC) |
|  | WC | WNC | WC | WNC |  |
| United States | 15503 | 7283.6 | 1 | 1 | 2.13 |
| China | 8526 | 4031.88 | 2 | 2 | 2.11 |
| United Kingdom | 8469 | 3984.37 | 3 | 3 | 2.13 |
| France | 7497 | 3566.65 | 4 | 4 | 2.10 |
| Germany | 6586 | 2999.97 | 5 | 5 | 2.20 |
| Japan | 4167 | 1904.52 | 6 | 6 | 2.19 |
| Switzerland | 3538 | 1650.27 | 7 | 7 | 2.14 |
| Canada | 3130 | 1416.02 | 8 | 8 | 2.21 |
| Russian Federation | 2917 | 1346.45 | 9 | 9 | 2.17 |
| Spain | 2732 | 1257.98 | 10 | 10 | 2.17 |
| Italy | 2085 | 948.17 | 11 | 11 | 2.20 |
| **Sweden** | 1905 | 859.35 | **12** | **13** | 2.22 |
| **India** | 1845 | 861.38 | **13** | **12** | 2.14 |
| Austria | 1570 | 735.07 | 14 | 14 | 2.14 |
| **Poland** | 1566 | 692.2 | **15** | **16** | 2.26 |
| **South Korea** | 1517 | 721.5 | **16** | **15** | 2.10 |
| Belgium | 1517 | 684.07 | 17 | 17 | 2.22 |
| **Netherlands** | 1414 | 597.3 | **18** | **19** | 2.37 |
| **Australia** | 1369 | 627 | **19** | **18** | 2.18 |
| Portugal | 1267 | 564.42 | 20 | 20 | 2.24 |
| Hong Kong | 1071 | 524.83 | 21 | 21 | 2.04 |
| Brazil | 1071 | 524.83 | 22 | 22 | 2.04 |

**CPP and ranking of top countries**

Table 3 shows the citation impact and relative rank of top countries by the two counting methods. The ranks for top 6 countries are similar by both the methods. Beyond this a fluctuation in ranks is observed. Of the 22 countries, there is a slightly change in the rank for 11 countries (50%). Inflation rate is calculated by dividing the CPP by whole counting with CPP by whole-normalized counting. It is observed that inflation rate of CPP from whole-normalized counting to whole counting of top countries ranges between 0.97 (Switzerland) and 1.09 (Netherlands). This inflation rate is in agreement withprevious studies in paper count where it was between 1.00 and 1.35 (Huang, Lin and Chen 2011)[2] and in patent count it was between 0.99 and 1.12 (Zheng, et al 2014)[8]. It is noticed that inflation rate for top 5 countries was lower than 1. This implies that the whole counting method underrated the top 5 country's impact of research (CPP).

| Table 3 – CPP and rankings | | | | | |
|---|---|---|---|---|---|
| Country | CPP | | Rank | | Inflation Rate (WC / WNC) |
| | WC | WNC | WC | WNC | |
| Switzerland | 19.44 | 20.03 | 1 | 1 | 0.97 |
| Russian Federation | 16.12 | 16.41 | 2 | 2 | 0.98 |
| Sweden | 15.24 | 15.44 | 3 | 3 | 0.99 |
| United Kingdom | 14.65 | 14.78 | 4 | 4 | 0.99 |
| France | 14.36 | 14.56 | 5 | 5 | 0.99 |
| United States | 14.12 | 14.02 | 6 | 6 | 1.01 |
| **Netherlands** | **13.21** | **12.12** | **7** | **8** | **1.09** |
| **Canada** | **12.62** | **12.38** | **8** | **7** | **1.02** |
| **Belgium** | **12.04** | **11.97** | **9** | **10** | **1.01** |
| **Spain** | **11.93** | **11.99** | **10** | **9** | **1.00** |
| **China** | **11.86** | **11.58** | **11** | **12** | **1.02** |
| **Austria** | **11.54** | **11.79** | **12** | **11** | **0.98** |
| **Portugal** | **11.31** | **11.12** | **13** | **14** | **1.02** |
| **Italy** | **11.27** | **11.19** | **14** | **13** | **1.01** |
| Germany | 11.09 | 10.82 | 15 | 15 | 1.02 |
| Australia | 10.22 | 10.05 | 16 | 16 | 1.02 |
| Japan | 9.67 | 9.55 | 17 | 17 | 1.01 |
| **Brazil** | **9.31** | **8.97** | **18** | **19** | **1.04** |
| **India** | **9.18** | **8.98** | **19** | **18** | **1.02** |
| Hong Kong | 8.85 | 8.97 | 20 | 20 | 0.99 |

| Poland | 8.33 | 7.93 | 21 | 21 | 1.05 |
| South Korea | 7.44 | 7.36 | 22 | 22 | 1.01 |

**h-index and ranking of top countries**

Table 4 show that the h-index and relative rank of top countries by two counting methods. The ranks for top countries are similar by two counting methods except eight countries where a slightly change is observed. Inflation rate is calculated by dividing the h-index by whole counting with h-index by whole-normalized counting. It is observed that inflation rate of h-index from whole-normalized counting to whole counting for top countries ranges between 1.26 (Hong Kong) and 1.37 (Netherlands).

| Table 4 – h-index and rankings | | | | | |
|---|---|---|---|---|---|
| Country | h-index | | Rank | | Inflation Rate (WC / WNC) |
| | WC | WNC | WC | WNC | |
| United States | 60.27 | 46.74 | 1 | 1 | 1.29 |
| United Kingdom | 49.88 | 38.91 | 2 | 2 | 1.28 |
| France | 47.57 | 37.31 | 3 | 3 | 1.28 |
| China | 46.59 | 36.01 | 4 | 4 | 1.29 |
| **Germany** | 41.80 | 31.90 | **5** | **6** | 1.31 |
| **Switzerland** | 40.97 | 32.09 | **6** | **5** | 1.28 |
| Russian Federation | 36.09 | 28.06 | 7 | 7 | 1.29 |
| Japan | 34.28 | 26.30 | 8 | 8 | 1.30 |
| Canada | 34.06 | 25.98 | 9 | 9 | 1.31 |
| Spain | 31.94 | 24.71 | 10 | 10 | 1.29 |
| Sweden | 30.73 | 23.67 | 11 | 11 | 1.30 |
| Italy | 28.64 | 21.98 | 12 | 12 | 1.30 |
| **Netherlands** | 26.54 | 19.35 | **13** | **16** | 1.37 |
| Belgium | 26.34 | 20.15 | 14 | 14 | 1.31 |
| **Austria** | 26.27 | 20.54 | **15** | **13** | 1.28 |
| **India** | 25.68 | 19.78 | **16** | **15** | 1.30 |
| **Portugal** | 24.29 | 18.45 | **17** | **18** | 1.32 |
| **Australia** | 24.09 | 18.47 | **18** | **17** | 1.30 |
| Poland | 23.54 | 17.64 | 19 | 19 | 1.33 |
| South Korea | 22.43 | 17.44 | 20 | 20 | 1.29 |
| Brazil | 21.53 | 16.76 | 21 | 21 | 1.28 |
| Hong Kong | 21.16 | 16.76 | 22 | 22 | 1.26 |

**Inflation rates**

Table 5 provides the information about lowest and highest inflation rates from whole-normalized countingto whole counting for various indicators. It is observed from table 5 that inflation rates for paper and citation counts from whole-normalized counting to whole counting are greater than 2 whereas it is greater than 1 for impact of research (CPP and h-index). This is because of counting done multiple times of countries for internationally collaborative papers.

| Table 5 – Inflation ratesbetween WC and WNC | | | |
|---|---|---|---|
| Indicator | Lowest Inflation Rate | Highest Inflation Rate | Average Inflation Rate |
| Paper count | 1.97 | 2.25 | 2.15 |
| Citation count | 2.04 | 2.37 | 2.17 |
| CPP | 0.97 | 1.09 | 1.01 |
| h-index | 1.26 | 1.37 | 1.30 |

**Summary of statistical tests**

Table 6 provides an overview of statistical tests performed in this study. Results of Pearsoncorrelation and Spearman's rank correlation tests show that all the indicators (paper count, citation count, CPP and h-index) between the two counting methods are highly correlated (>0.990 at the 0.01 significance level). Further results analyzed using independent samples t-test reveals that there is significant difference in paper count, citation count and h-index except CPP where there is no significant difference between the two counting methods.

| Table 6 –Statistical tests on various indicators between WC and WNC | | | |
|---|---|---|---|
| Indicator | Pearson | Spearman | T-test |
| Paper count | 1 (0.00) | 0.990 (0.00) | t = 2.611 (p = 0.01246)** |
| Citation count | 1 (0.00) | 0.996 (0.00) | t = 2.344 (p = 0.02385)** |
| CPP | 0.995 (0.00) | 0.995 (0.00) | t = 0.092 (p = 0.92743) |
| h-index | 0.999 (0.00) | 0.990 (0.00) | t = 2.589 (p = 0.01316)** |
| **Statistically significant difference at the p < 0.05 level | | | |

**CONCLUSION**

This study compares the two counting methods namely whole counting and whole-normalized counting based on the research productivity and impact at country level on tribology research as basic data. The result of Pearson and spearman's test indicates that the choice of counting method does not affect greatly the country rankings. However the results of t-test show that that there is significant difference in paper count, citation count and h-index among the two counting methods except CPP. The distributions of higher and lower inflation rate by the two counting methods are different in all the four indicators such as, paper count, citation count, CPP and h-index (Zheng, et al 2014)[8]. The difference between the lowest and highest inflation rates is very low among the two counting methods, i.e. from 0.11 for h-index to 0.33 for citation count. This result is in contrast to previous study conducted by Huang, et al (2011)[2] where it was 0.63 for paper count, 0.92 for citation count and 0.35 for CPP. The inflation rate was lower than 1 in CPP for seven countries (32% of top 22 countries). Even though correlation coefficient between the two counting methods is very close to 1.0, the inflation rate is greater than 2 for paper and citation counts whereas it is greater than 1 for impact of research (CPP and h-index). Considering this high inflation rate, this study concludes that whole-normalized counting (fractional counting) is better choice than whole counting at the national level research evaluation. This result is in agreement with the argument by Waltman, et al (2012)[5] and Aksnes, et al (2012)[7].

There are three limitations in this study. A relatively small dataset is used for comparison. Secondly, simple fractional counting (whole-normalized counting) is used rather than complete-normalized counting, as one unit of credit is shared between the unique countries with equal fractions (OECD 2009)[17]. Thirdly, the number of citations received by a paper from its time of publication through the date of access is used rather than specific period citation windows such as two year citations.

**ACKNOWLEDGMENTS**

The authors are thankful to Prof. Isola Ajiferuke (Faculty of Information and Media Studies, University of Western Ontario, Canada) for providing useful comments and suggestions. The authors also thank to Prof. Matilda S (Vice Principal and Dean, IFET College of Engineering, Villupuram, India) who helped us proof reading the manuscript.

**REFERENCES**

1. Huang M H and Lin C S (2011). Counting methods and university rankings by H-index. *Proceedings of the American Society of Information Science and Technology*, 48 (1): 1-6.

2. Huang M H, Lin C S and Chen D Z (2011). Counting methods, country rank changes, and counting inflation in the assessment of national research productivity and impact. *Journal of the American Society for Information Science and Technology*, 62 (12): 2427-2436.

3. Lin C S, Huang M H and Chen D Z (2013). The influences of counting methods on university rankings based on paper count and citation count. *Journal of Informetrics*, 7 (3): 611-621.

4. Egghe L, Rousseau R and van Hooydonk G (2000). Methods for accrediting publications to authors or countries: Consequences for evaluation studies. *Journal of the American Society of Information Science*, 51 (2): 145-157.

5. Waltman L, et al (2012). The Leiden ranking 2011/2012: Data collection, indicators and interpretation. *Journal of the American Society for Information Science and Technology*, 63 (12): 2419-2432.

6. Chudlarský T, Dvořák J and Souček M (2014). A comparison of research output counting methods using a national CRIS − effects at the institutional level. *Procedia Computer Science*, 33: 147-152.

7. Aksnes D W, Schneider J W and Gunnarsson M (2012). Ranking national research systems by citation indicators: A comparative analysis using whole and fractionalized counting methods. *Journal of Informetrics*, 6 (1): 36-43.

8. Zheng J, et al (2014). Influences of counting methods on country rankings: a perspective from patent analysis. *Scientometrics*, 98: 2087-2102.

9. Huang M H and Lin C S (2010). International collaboration and counting inflation in the assessment of national research productivity. *Proceedings of the American Society of Information Science and Technology*, 47 (1): 1-4.

10. Gauffriau M, et al (2008). Comparisons of results of publication counting using different methods. *Scientometrics*, 77 (1): 147-176.

11. Sung S Y, et al (2014). A comparative study of patent counts by the inventor country and the assignee country. *Scientometrics*, 100 (2): 577-593.

12. Elango B, Rajendran P and Bornmann L (submitted). A macro level scientometric analysis of world tribology research output (1998-2012). *Malaysian Journal of Library and Information Science.*

13. Elango B, Rajendran P and Bornmann L (2013). Global nanotribology research output (1996-2010): A scientometric analysis. *PLOS ONE*, 8 (12): e81094.

14. Hirsch J E (2005). An index to quantify an individual's scientific research output. *Proceedings of the National Academy of Sciences*, 102 (46): 16569-16572.

15. Schubert A (2007). Successive h-indices. *Scientometrics*, 70 (1): 201-205.

16. Schubert A and Glänzel W (2007). A systematic analysis of Hirsch-type indices for journals. *Journal of Informetrics*, 1 (2): 179-184.

17. OECD (2009). *OECD Patent statistics manual*. Paris : OECD.